\documentclass[runningheads]{llncs}
\usepackage{hyperref}
\usepackage{booktabs} 
\usepackage{latexsym}
\usepackage{graphicx}
\usepackage{multirow}
\usepackage{url}
\usepackage{subcaption}
\usepackage{amsmath}
\usepackage{amsfonts}

\DeclareMathOperator*{\argmax}{argmax}

\begin{document}
\title{Modeling Information Need of Users in \\Search Sessions}

\author{Kishaloy Halder\inst{1}\thanks{Work done during internship at Google Research.} \and
Heng-Tze Cheng\inst{2} \and
Ellie Ka In Chio\inst{2} \and
Georgios Roumpos\inst{2} \and \\
Tao Wu\inst{2} \and
Ritesh Agarwal\inst{2}}
\authorrunning{K. Halder et al.}
%
\institute{School of Computing, National University of Singapore\\
\email{kishaloy@comp.nus.edu.sg}\\ \and
Google Research\\
\email{\{hengtze,echio,roumposg,iotao,riteshag\}@google.com}}

\maketitle

\begin{abstract}
 Users issue queries to Search Engines, and try to find the desired information in the results produced. They repeat this process if their information need is not met at the first place.
It is crucial to identify the important words in a query that depict the actual information need of the user and will determine the course of a search session.
To this end, we propose a sequence-to-sequence based neural architecture that leverages the set of past queries issued by users, and results that were explored by them.
Firstly, we employ our model for 
predicting the words in the current query that are important and would be retained in the next query. 
Additionally, as a downstream application of our model, we evaluate it on the widely popular
task of next query suggestion. 
We show that our intuitive strategy of capturing information need can yield superior performance at these tasks on two large real-world search log datasets.
\end{abstract}

\section{Introduction}
\begin{figure}[t!]
\centering
\includegraphics[width=0.9\textwidth]{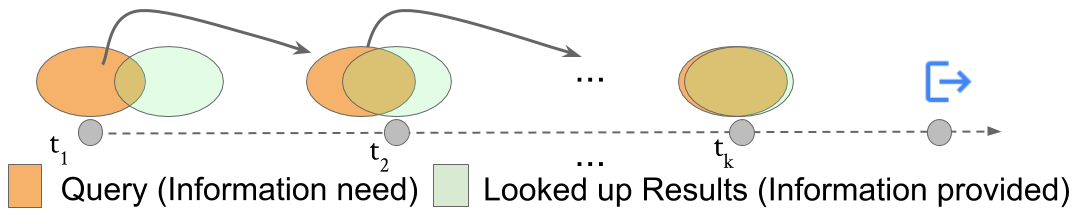}

\caption{Illustration of the user-search engine interactions. User fires queries at different timestamps depending on the ``gap'' between the information need (orange), and information provided (green). User ends the session when orange $\approx$ green.}
\label{figure:timeline}
\end{figure}

In a search session, if a user's need is not completely fulfilled by the results returned by the Search Engine (SE), they tend to modify their query, and ask again---leaving a recurring footprint of the nature \textit{query $\rightarrow$ clicks $\rightarrow$ query $\rightarrow$ ... } in the logs. 
This set of interactions can be viewed as an exchange of information between two agents i.e., user and SE (Figure \ref{figure:timeline}). The user has an information need which is reflected in the query issued by her (at time $t_1$). The results returned by the SE provide some (but may not be all) information relevant to the query. We hypothesize that in this situation, the next query (at time $t_2$) would be dependent on the information deficit that has occurred so far, \textit{i.e.,} the semantic gap between the information need, and the information provided.

\begin{figure}[t]
\centering
\includegraphics[width=1\textwidth]{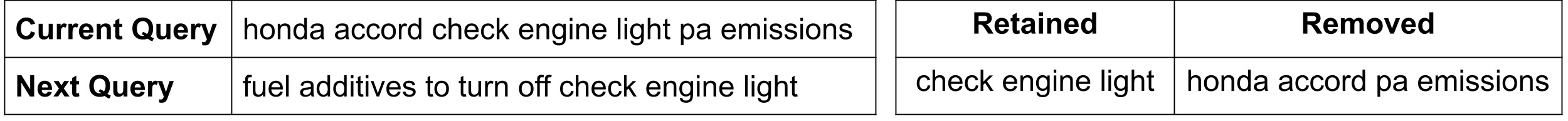}
\caption{Illustration of query edit from AOL. Table on the right is the prediction target.}
\label{figure:sample}
\end{figure}

Important words play a crucial role in depicting the real information need behind a search.
Consider an illustration shown in Figure \ref{figure:sample}. Users tend to retain the important words in the subsequent queries (\textit{e.g.,} `check', `engine', `light'), and remove the ones that are not (\textit{e.g.,} `honda', `accord' etc). To model this behavior, we aim to answer the following research question (RQ).

{\textbf{RQ:}} Can the information deficit help in predicting the query \textit{edits}?

We define query edits by means of two possibilities for every word in it \textit{i.e.,} user may (i) retain or (ii) remove a word in the next query. 
To answer this, in this work we address the task of
estimating the probability of retention for every word in the current query, given the past interactions within a session. 
We believe that being able to predict which words would be dropped later in the session 
(`honda', `accord' in the example shown in Figure \ref{figure:sample}) 
can help a search engine focus on the core issue 
(`check', `engine', `light') 
and diversify its search results. This will help  meet a user's information need better and faster.
We propose a novel neural encoder-decoder architecture 
which explores the role of evolving information deficit experienced by the user in a search session. By experimenting with two large real-world query log datasets from AOL, and Yandex we show that the information deficit does help in predicting the query edits. 

As a downstream application, we employ our model to the widely researched next query selection task. We observe that it outperforms state-of-the-art baselines significantly, indicating the untapped potential of the evolving information need present in search sessions.

\section{Related Work}
\label{sec:rel_work}
Query word deletion prediction has been seen as a strategy for query rewriting in the past \cite{jones2003query,wu2011exploration}. They propose some hand-coded rules such as deletions from ``leftmost'', ``rightmost'' positions in the search query. Although, these methods do not capture the query context effectively \cite{yang2014study}, they present some insights about the nature of query modifications. Another interesting direction is reducing the verbosity of queries \cite{balasubramanian2010exploring,kumaran2009reducing,bendersky2008discovering,huston2010evaluating,xue2010improving}.
These methods in general lack the context of past queries in a session where users' information needs keep evolving after exploring some of the result pages.

Next query suggestion, and query reformulation behaviour modeling is a heavily researched area. Early works in this domain presented the concept of semantic similarity by looking at the queries co-occurring within a session \cite{boldi2008query}, query clustering by utilizing the query-click graph \cite{jain2011synthesizing}. 
Next query formulation is also seen as a generative task  \cite{he2009web,cao2008context}. 
Unsurprisingly, the recent deep learning based methods outperform previous heuristic based approaches. Introduced in \cite{sordoni2015hierarchical}, Hierarchical Recurrent Encoder Decoder (HRED) uses two-level recurrent neural network (RNN) to first consider the sequence of words in a query, and the sequence of past queries to make the prediction. Attention based networks were also proposed which focuses on the out of vocabulary words in the query \cite{dehghani2017learning}. To incorporate user feedbacks, use of memory network was proposed recently \cite{wu2018query}. Although these models \textit{are} context aware, they do not explore the relationship between the different contextual signals which are indicative of the users' information need.
In our experiments we have compared our model to these systems.

\section{Methods}
\label{sec:methods}
\begin{figure*}[t]
    \centering
    \begin{subfigure}[b]{0.5\textwidth}
         \centering
        \includegraphics[scale=0.25]{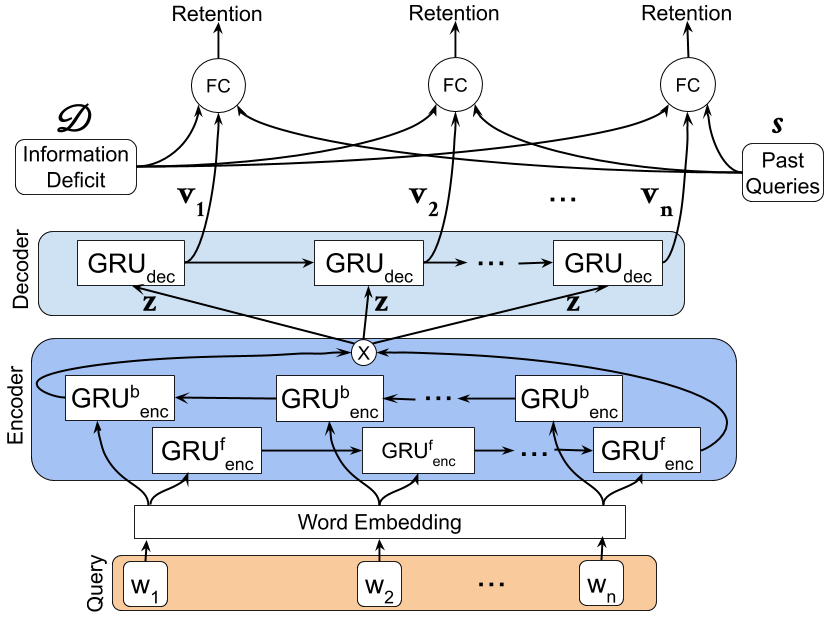}
        \caption{Overall architecture.}
        \label{figure:full_model}
     \end{subfigure}%
     \hfill
     \begin{subfigure}[b]{0.5\textwidth}
         \centering
        \includegraphics[scale=0.20]{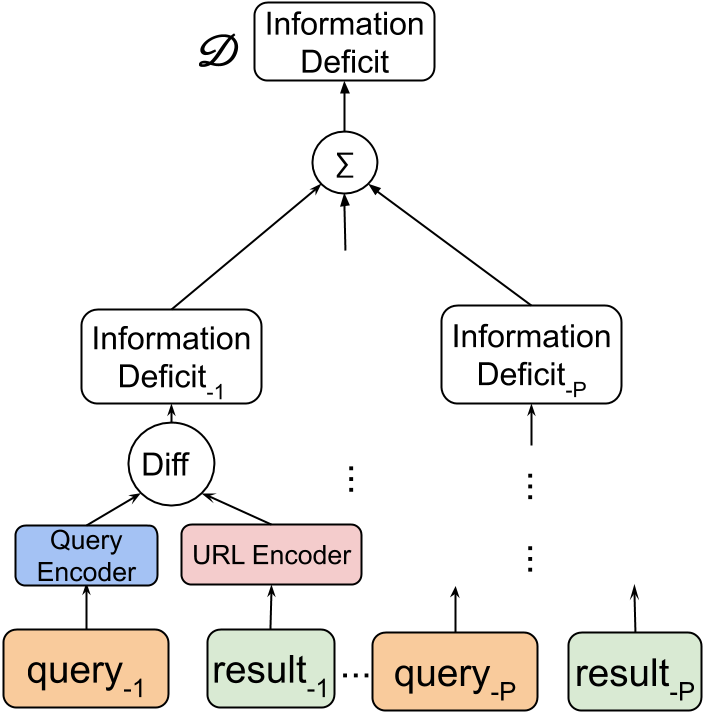}
        \caption{Information Deficit.}
        \label{figure:info_deficit}
     \end{subfigure}%
    \caption{Our model architecture with the Information Deficit component. (a) Overall network. (b) Information Deficit for the current query. We consider the sum of the past $P$ Information Deficits w.r.t a query as the final one.
    }
    \label{figure:architecture}
\end{figure*}

Our model comprises two
major components \textit{i.e.,} (i) Query Encoder, and (ii) Information Deficit Encoder as shown in Figure \ref{figure:architecture}.

\begin{figure*}[t]
    \centering
    \begin{subfigure}[b]{0.5\textwidth}
         \centering
        \includegraphics[scale=0.25]{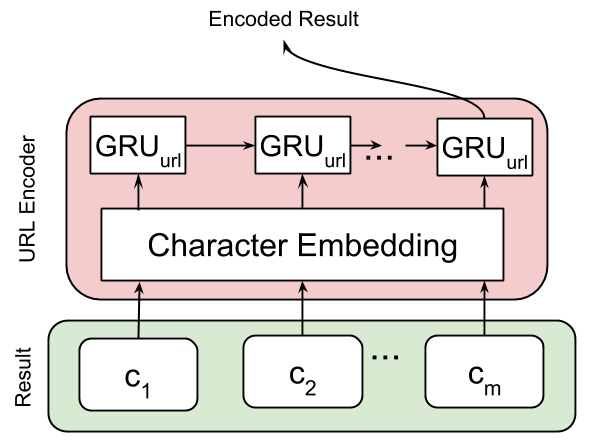}
        \caption{URL encoder}
        \label{figure:url_encoder}
     \end{subfigure}%
    \hfill
     \begin{subfigure}[b]{0.5\textwidth}
         \centering
        \includegraphics[scale=0.15]{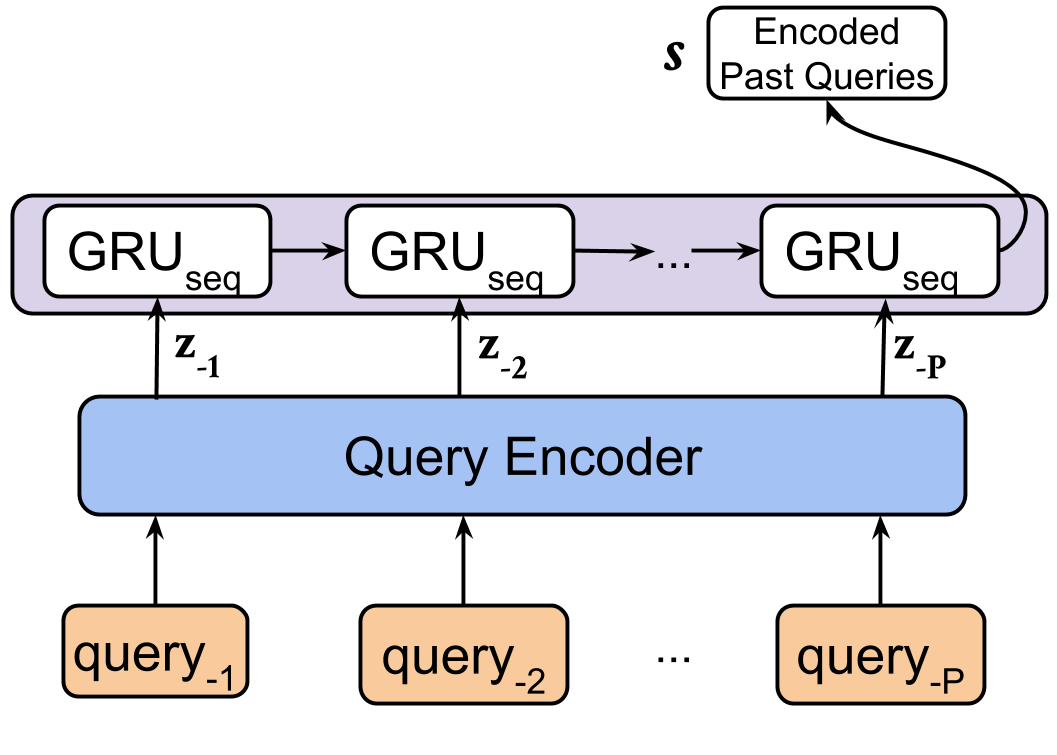}
        \caption{Encoding Sequence of Past Queries.}
        \label{figure:seq_encoder}
     \end{subfigure}%
    \caption{Components in our network. (a) Architecture diagram for URL encoder. A Character Embedding layer is used for all the characters in a URL. $\text{GRU}_\text{url}$ encodes the sequence of characters present in it. (b) Architecture for encoding sequence of past queries. Output of last time-step of $\text{GRU}_\text{seq}$ is considered as the encoding of past queries.}
    \label{figure:components}
\end{figure*}

\noindent{\textbf{Query Encoder.}} 
Given a query $q$ with $n$ words, we use a bi-directional Gated Recurrent Unit to encode all the words in it \cite{chung2014}. 
Following are the set of equations used:
\begin{equation}
\mathbf{h}_{i}^{f} = \mathbf{\text{GRU}}_{\text{enc}}^{f}(w_i, \mathbf{h}_{i-1}^{f}); \quad \mathbf{h}_{i}^{b} = \mathbf{\text{GRU}}_{\text{enc}}^{b}(w_i, \mathbf{h}_{i+1}^{b}); \quad
\mathbf{z} = \mathbf{h}_{n}^{f} \otimes \mathbf{h}_{1}^{b}
\label{equation:encoder}
\end{equation}
where, $w_i$ is the $i^{th}$ query word, $\mathbf{h}_{n}^{f}$, $\mathbf{h}_{1}^{b}$ are the GRU encoded representations after the final 
(i.e., $n^{\text{th}}$ and $1^{\text{st}}$) time-steps for forward, and backward passes respectively, 
$\mathbf{z}$ is the final encoded representation of the query, $\otimes$ denotes element-wise multiplication operation.
The encoded query($\mathbf{z}$) is repeated $n$ times, and is fed to $\text{GRU}_\text{dec}$ as a sequence. $\text{GRU}_\text{dec}$ yields a tensor($\mathbf{v}_i$) for each time-step to be used for final prediction later.\\

\noindent{\textbf{Information Deficit Encoder.}}
We carefully encode two contextual signals (i) $P$ past queries issued by a user, and (ii) the resources she has seen in the session so far. We combine these two signals to compute information deficit. We encode each of the past queries using the same Query Encoder 
as in Equation \ref{equation:encoder}. We denote the encoded representation of $j^{\text{th}}$ past query $q_{-j}$ w.r.t the current query $q$ as, $\mathbf{z_{-j}}$.

After issuing a query, a user tries to find the desired information in the  results and modifies her query accordingly. Ideally, content of the landing pages should be considered as result. However, those were not available\footnote{Most of the urls from AOL log do not exist anymore. Yandex urls are completely anonymized.}.
We encode the results using the component in Figure \ref{figure:url_encoder}. We concatenate all the urls (in the original order) clicked against a query and consider that as a proxy for the result.
Since urls are often comprising non-standard tokens, they form a highly sparse embedding space. To alleviate this, we consider result $r$ as a sequence of characters $\{c_1, c_2, ..., c_m\}$. We use $\text{GRU}_\text{url}$ which considers 
a character at every time-step, and output of the last time-step is considered as the encoded result i.e., $\mathbf{u} = \mathbf{h^\text{url}_m}$ where $\mathbf{h^\text{url}}_{k} = \text{GRU}_\text{url}(c_{k}, \mathbf{h^\text{url}}_{k-1}); \mathbf{h^\text{url}}_{k}$ is output after $k^{\text{th}}$ character in the result.
Once we have the encoded representations $\mathbf{z_{-j}}$ for the $j^{\text{th}}$ past query, and $\mathbf{u_{-j}}$ for the results explored against it, we define the information deficit $(\mathbf{\mathcal{D}_{-j}})$ as, 

\[\mathbf{\mathcal{D}_{-j}} = \mathbf{z}_{-j} \ominus \mathbf{u}_{-j}\]

where, $\ominus$ denotes an element-wise subtraction operation. Finally, we sum up the information deficits for all the $P$ past queries to get the overall information deficit, 
\[\mathbf{\mathcal{D}} = \sum_{j=1}^P\mathbf{\mathcal{D}_{-j}}\]

\noindent{\textbf{Final Prediction Layer.}}
We combine the (i) current query representation, (ii) overall information deficit and (iii) an encoded representation of the past queries to make the final prediction. We use $\text{GRU}_\text{seq}$ to encode the sequence of past query representations $\mathbf{z}_{-1},...,\mathbf{z}_{-P}$ and denote it as $\mathbf{s}$ (shown in Figure \ref{figure:seq_encoder}). Since the final prediction is a binary label for each word in the query, we repeat the information deficit and past query sequence tensors $n$ times.
For $i^{\text{th}}$ word, we concatenate these three tensors, and pass the resultant tensor 
$\mathbf{x}_{i}$ to the fully connected(FC) layers to make the binary prediction.
\[\mathbf{x}_{i} = \text{concatenate}[\mathbf{v}_{i}, \mathbf{\mathcal{D}}, \mathbf{s}]; \quad p(y_{i}|\mathbf{x}_{i}) = softmax(\mathbf{W} \cdot \mathbf{x}_{i} + \mathbf{b})\]
where $\mathbf{W}$, and $\mathbf{b}$ are the weight matrices, and bias vector, 
respectively for the fully connected layer; and $y_{i}$ is a probability distribution over two prediction classes i.e., \textit{retention}, \textit{removal} for the $i^{\text{th}}$ word.
The loss is \textit{binary cross-entropy} and the network is trained end-to-end using \textit{Adam} optimizer \cite{kingmaB14}.
\vspace{-1em}

\section{Experiments}
We experiment with two large datasets: (i) AOL: a US-based search engine, widely used in the literature \cite{dehghani2017learning}; (ii) Yandex\footnote{\url{https://www.kaggle.com/c/yandex-personalized-web-search-challenge}}: a search engine based in Russia.\\

\noindent{\textbf{Pre-processing:}} To focus on the query reformulation behaviour we consider only those queries that (i) are not the same as the subsequent query (ii) have more than a single word (iii) are not navigational in nature i.e. do not include http markers such as `www', `com' (iv) have at least one word in common with the next query. 
We split all the queries (AOL $\sim3.6M$ queries from $1.5M$ sessions, Yandex $\sim11M$ queries from $6.6M$ sessions) into 80-10-10 proportions as train-dev-test. We kept top 600K, 400K words as the vocabulary for AOL, and Yandex respectively. Since Yandex urls are just anonymized ids, only top 200K urls are kept.\\

\noindent{\textbf{Baselines:}} We use some recent deep learning based query reformulation methods as competitors, Since they are not originally designed for this task, we modify their final prediction layers for comparison.

\noindent{\textbf{1. Majority Class:}} Every word is marked `retention' forming a trivial baseline.

\noindent{\textbf{2. CNN-Kim \cite{kim2014convolutional}:}} A CNN without the pooling layer to predict a label for each word.

\noindent{\textbf{3. GRU \cite{chung2014}:}} A vanilla GRU based sequence labeller.

\noindent{\textbf{4. HRED \cite{sordoni2015hierarchical}:}} The sequence of past queries in the same session is used in this model.

\noindent{\textbf{5. FMN \cite{wu2018query}:}} A feedback memory network is used in this model.

The baseline hyper-parameters are set according to the respective papers. All models were implemented using Keras with TensorFlow as the backend.

\noindent{\textbf{Word Retention Prediction:}}
In this task, the objective is to predict how a user would edit a query. 
Since there are two classes \textit{i.e.,} `removal', and `retention', we calculate accuracy over all the words in a query, normalized by the query length.
We report the performances in terms of overall accuracy, and $\text{F}_1$ score for the `removal' class in Table \ref{table:scores}, since the class distribution is not balanced.
Table \ref{table:accuracy} shows an illustration.
\vspace{-1cm}
\begin{table}
\centering
\caption{Illustration of accuracy computation for query word retention probability prediction task.}
\label{table:accuracy}
\resizebox{0.9\textwidth}{!}{%
\begin{tabular}{|c|c|c|c|c|}
\hline
Current Query & Next Query & \begin{tabular}[c]{@{}c@{}}Ground \\ Truth\end{tabular} & Prediction & Accuracy \\ \hline
\textit{japanese} food for takeout & asian food for takeout & 0, 1, 1, 1 & 0, 1, 1, 0 & 0.75 \\ \hline
cheap electronics \textit{bay area} & cheap electronics offers & 1, 1, 0, 0 & 0, 1, 0, 1 & 0.5 \\ \hline
\multicolumn{4}{|c|}{Average Accuracy} & 0.625 \\ \hline
\end{tabular}%
}
\vspace{-0.5cm}
\end{table}
The accuracy scores are presented in Table \ref{table:scores}. 
We observe that our information deficit based model achieves the highest accuracy scores with $9.22\%$ and $5.88\%$ improvement over the majority class for the AOL, and Yandex respectively. 
\begin{table}[t]
\centering
\caption{Experimental results in terms of accuracy and $\text{F}_1$ scores
for the removal class. Our model outperforms the other methods in both metrics.}
\label{table:scores}
\resizebox{0.9\textwidth}{!}{%
\begin{tabular}{|l|c|c|c|c|c|c|c|}
\hline
\multicolumn{1}{|c|}{\multirow{2}{*}{Model}} &
\multicolumn{3}{c|}{Input Signals} &
\multicolumn{2}{c|}{AOL} & \multicolumn{2}{c|}{Yandex} \\ \cline{2-8} 
\multicolumn{1}{|c|}{} & Last Query & Past Queries & Past Clicks & Accuracy & $\text{F}_1$ & Accuracy & $\text{F}_1$ \\ \hline
1. Most Common & $\times$ & $\times$ & $\times$ & $.628$ & -- & $.676$ & -- \\ \hline
2. CNN-Kim \cite{kim2014convolutional}& $\checkmark$ & $\times$ & $\times$ & $.664 \pm .0004$ & $.36$ & $.690 \pm .0086$ & $.36$ \\ \hline
3. GRU \cite{chung2014}& $\checkmark$ & $\times$ & $\times$ & $.678 \pm .0010$ & $.42$ & $.711 \pm .0007$ & $.40$ \\ \hline
4. HRED \cite{sordoni2015hierarchical}& $\checkmark$ & $\checkmark$ & $\times$ & $.684 \pm .0011$ & $.44$ & $.713 \pm .0007$ & $.41$ \\ \hline
5. FMN \cite{wu2018query}& $\checkmark$ & $\checkmark$ & $\checkmark$ & $.683 \pm .0005$ & $.45$ & $.714 \pm .0008$ & $.41$ \\ \hline \hline
6. Our Model & $\checkmark$ & $\checkmark$ & $\checkmark$ & $\mathbf{.686^*} \pm .0006$ & $\mathbf{.47^*}$ & $\mathbf{.716^*} \pm .0005$ & $\mathbf{.43^*}$ \\ \hline
\end{tabular}
}
\vspace{-0.5cm}
\end{table}

Our model performs significantly better in correctly identifying the removal class with $4.45\%$ and $4.87\%$ improvement over the state-of-the-art FMN model. The choice of baselines allows us to study the ablation effect. We observe that information deficit helps in boosting the $\text{F}_1$ scores by $6.81\%$ and $4.87\%$ w.r.t HRED for AOL, and Yandex respectively. The improvement with respect to FMN is statistically significant according to unpaired t-test with $p<0.001$.

A careful reader might note that both FMN and our model use all three input signals. However, the information deficit based modelling makes a difference their performances. FMN incorporates the past clicks as a separate signal compared to the past queries. Whereas, our model assumes that both originate from the same semantic space and uses their difference in the final prediction layers. It proves our hypothesis about encoding information deficit explicitly in the query reformulation process.\\

\noindent{\textbf{Next Query Selection:}}
Finally, we employ our model to a real downstream task \textit{i.e.,} next query selection in a session. 
The objective in this task is to select the next query out of $K$ most popular candidates. For each query under consideration, we populate the top-K $(K=20)$ co-occurred queries that to form the candidate set, and include the ground truth next query. This is known as Most Popular Suggestion method \cite{sordoni2015hierarchical}. Standard 80-10-10 dataset splitting is used for train-dev-test.

\begin{figure*}[t]
    \centering
    \begin{subfigure}[b]{0.5\textwidth}
         \centering
        \includegraphics[scale=0.25]{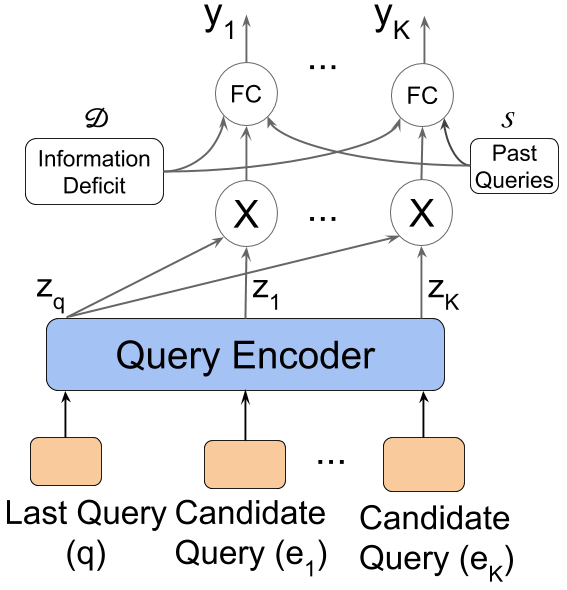}
        \caption{Model  for Query Selection Task.}
        \label{figure:query_selection}
     \end{subfigure}%
     \hfill
     \begin{subfigure}[b]{0.5\textwidth}
            \resizebox{1\textwidth}{!}{%
            \begin{tabular}{|l|c|c|c|c|c|}
            \hline
            \multicolumn{1}{|c|}{\multirow{2}{*}{Model}} & \multicolumn{3}{|c|}{Input Signals} & \multicolumn{2}{c|}{MRR} \\ \cline{2-6} 
            \multicolumn{1}{|c|}{} & LQ & PQ & PC & AOL & Yandex \\ \hline
            2. CNN-Kim \cite{kim2014convolutional}& $\checkmark$ & $\times$ & $\times$ & $.752$ & $.621$ \\ \hline
            3. GRU \cite{chung2014} & $\checkmark$ & $\times$ & $\times$ & .773 & .636 \\ \hline
            4. HRED \cite{sordoni2015hierarchical} & $\checkmark$ & $\checkmark$ & $\times$ & $.814$ & $.674$ \\ \hline
            5. FMN \cite{wu2018query} & $\checkmark$ & $\checkmark$ & $\checkmark$ & $.853$ & $.686$ \\ \hline \hline
            6. Our Model & $\checkmark$ & $\checkmark$ & $\checkmark$ & $\mathbf{.864*}$ & $\mathbf{.693*}$ \\ \hline
            \end{tabular}%
            }
            \caption{MRR scores for the competing methods in the next query suggestion task. Our model outperforms the baseline methods.}
            \label{table:mrr}
     \end{subfigure}%
     \vspace{-0.3cm}
    \caption{Next Query Selection Setup. (a) Modified architecture (b) Experimental Results.
    }
    \label{figure:architecture2}
\end{figure*}

This task is considerably different from the previous one. We modify the architecture as shown in Figure \ref{figure:query_selection}, keeping our core information deficit component unchanged. We encode the last query, and all the candidate queries using the same Query Encoder ({\it cf.} Figure \ref{figure:architecture}). 
Note that, we need a single representation of the entire query in this case. Hence, we consider output of the last ($n^{\text{th}}$) time-step of $\text{GRU}_\text{dec}$ as the query representation $\mathbf{z}_q$. 
\begin{equation}
    \mathbf{v}_i = \text{GRU}_\text{dec}(\mathbf{z}, \mathbf{v}_{i-1}); \quad \mathbf{z}_q = \mathbf{v}_n
\end{equation}
The candidate queries ($e_1, \cdots, e_K$) are encoded in the same manner ($\mathbf{z}_{1}, \mathbf{z}_2,\cdots, \mathbf{z}_k$). 
We perform element-wise multiplication between its encoded representation and the current query to obtain the similarity, i.e., $sim(q, e_j) = \mathbf{z}_q \otimes \mathbf{z}_j$. Again, we concatenate the information deficit, and past query tensor with each of the similarity tensors to obtain $\mathbf{x}_j = \text{concatenate}[sim(q, e_j), \mathcal{D}, \mathbf{s}]$.
These ($\mathbf{x}_j$) are fed to the fully-connected layer for the final score prediction.  
\[p(y_j|\mathbf{x}_{j}) = \text{sigmoid}(\mathbf{W'}\cdot \mathbf{x}_j + \mathbf{b'})\]
where $\mathbf{W'}$, and $\mathbf{b'}$ are the weight and bias matrices respectively. 
The ground truth score for the correct next query is $1$, and $0$ for the incorrect ones. We use $sigmoid$ as the activation function in final layer, and \textit{mean absolute error} as the loss. During test, we consider the candidate with the maximum score as the suggested next query, 
\[\hat{q} = \argmax_{j=\{1,\cdots,K\}} p(y_j|\mathbf{x}_j)\]

We evaluate the competing methods by computing Mean Reciprocal Rank (MRR) over test queries. From the scores in Table \ref{table:mrr}, we observe that our model consistently outperforms the baselines across the datasets. We find that the improvement with respect to FMN is statistically significant according to unpaired t-test with $p<0.001$. 
In this work, we focus on the exchange of information between a user and the Search Engine in a session to understand the information need of the user. Our model is based on the principle that a user forms a query depending on the information deficit that takes place during search sessions. Through a novel query word retention prediction task, and the traditional next query selection task, we show that our hypothesis generalizes well. We experiment with two query log datasets with considerably different demographic and temporal characteristics (e.g., one is from US, other from Russia; time difference of $8$ years between their releases) and show that our model works well in both cases. In the future, we would like to investigate on how to use such predictive models for converging a search session faster.

\bibliographystyle{splncs04}
\bibliography{sample-bibliography}

\end{document}